\documentclass[journal=apchd5,manuscript=article]{achemso}

\usepackage[T1]{fontenc} % Use modern font encodings
\usepackage{comment}
\usepackage{graphicx} %Loading the package
\graphicspath{{figures/}}
\usepackage{float}
\usepackage{amsmath} %For align environment
\usepackage[separate-uncertainty = true, multi-part-units=single, range-phrase=--, range-units = single, detect-all]{siunitx}
\author{Siqi Qiao}
\affiliation[Forschungszentrum Jülich GmbH]
{Peter Grünberg Institute (PGI-9), Forschungszentrum Jülich GmbH, 52428 Jülich, Germany}
\alsoaffiliation[JARA]
{JARA-Fundamentals of Future Information Technology, Jülich-Aachen Research Alliance, 52074 Aachen, Germany}
\email{s.qiao@fz-juelich.de}

\author{Nils von den Driesch}
\affiliation[Forschungszentrum Jülich GmbH]
{Peter Grünberg Institute (PGI-10), Forschungszentrum Jülich GmbH, 52428 Jülich, Germany}
\alsoaffiliation[JARA]
{JARA-Fundamentals of Future Information Technology, Jülich-Aachen Research Alliance, 52074 Aachen, Germany}

\author{Xi Chen}
\affiliation[MIT]
{Department of Materials Science and Engineering, Massachusetts Institute of Technology, Cambridge, MA 02139}

\author{Stefan Trellenkamp}
\affiliation[Forschungszentrum Jülich GmbH]
{Helmholtz Nano Facility (HNF), Forschungszentrum Jülich GmbH, 52428 Jülich, Germany}

\author{Florian Lentz}
\affiliation[Forschungszentrum Jülich GmbH]
{Helmholtz Nano Facility (HNF), Forschungszentrum Jülich GmbH, 52428 Jülich, Germany}

\author{Christoph Krause}
\affiliation[Forschungszentrum Jülich GmbH]
{Peter Grünberg Institute (PGI-10), Forschungszentrum Jülich GmbH, 52428 Jülich, Germany}

\author{Benjamin Bennemann}
\affiliation[Forschungszentrum Jülich GmbH]
{Peter Grünberg Institute (PGI-10), Forschungszentrum Jülich GmbH, 52428 Jülich, Germany}

\author{Thorsten Brazda}
\affiliation[Forschungszentrum Jülich GmbH]
{Peter Grünberg Institute (PGI-9), Forschungszentrum Jülich GmbH, 52428 Jülich, Germany}

\author{James M. LeBeau}
\affiliation[MIT]
{Department of Materials Science and Engineering, Massachusetts Institute of Technology, Cambridge, MA 02139}

\author{Alexander Pawlis}
\affiliation[Forschungszentrum Jülich GmbH]
{Peter Grünberg Institute (PGI-9), Forschungszentrum Jülich GmbH, 52428 Jülich, Germany}
\alsoaffiliation[Forschungszentrum Jülich GmbH]
{Peter Grünberg Institute (PGI-10), Forschungszentrum Jülich GmbH, 52428 Jülich, Germany}
\alsoaffiliation[JARA]
{JARA-Fundamentals of Future Information Technology, Jülich-Aachen Research Alliance, 52074 Aachen, Germany}
\email{a.pawlis@fz-juelich.de}

\title[paper]
  {Two-dimensional photonic crystal cavities in ZnSe quantum well structures}

\abbreviations{photoluminescence (PL), photonic crystal (PC), resonance scattering spectroscopy (RSS), single photon source (SPS)}

\begin{document}

\onehalfspacing
\newpage
%248 words, limit is 250 words.
\begin{abstract}
ZnSe and related materials like ZnMgSe and ZnCdSe are promising II-VI host materials for optically mediated quantum information technology such as single photon sources or spin qubits. Integrating these heterostructures into photonic crystal (PC) cavities enables further improvements, for example realizing Purcell-enhanced single photon sources with increased quantum efficiency. Here we report on the successful implementation of two-dimensional (2D) PC cavities in strained ZnSe quantum wells (QW) on top of a novel AlAs supporting layer. This approach overcomes typical obstacles associated with PC membrane fabrication in strained materials, such as cracks and strain relaxation in the corresponding devices. We demonstrate the attainment of the required mechanical stability in our PC devices, complete strain retainment and effective vertical optical confinement. Structural analysis of our PC cavities reveals excellent etching anisotropy. Additionally, elemental mapping in a scanning transmission electron microscope confirms the transformation of AlAs into AlOx by post-growth wet oxidation and reveals partial oxidation of ZnMgSe at the etched sidewalls in the PC. This knowledge is utilized to tailor FDTD simulations and to extract the ZnMgSe dispersion relation with small oxygen content. Optical characterization of the PC cavities with cross-polarized resonance scattering spectroscopy verifies the presence of cavity modes. The excellent agreement between simulation and measured cavity mode energies demonstrates wide tunability of the PC cavity and proves the pertinence of our model. This implementation of 2D PC cavities in the ZnSe material system establishes a solid foundation for future developments of ZnSe quantum devices.
\end{abstract}
%\begin{multicols}{2}
\newpage
\section{Introduction}
Photonic crystals (PC) structures, which are characterized by periodic variations in their refractive indices, possess the ability to modulate light generation and propagation by engineering the photonic band gap. Due to their small mode volume and moderately high Q-factor, PC cavities are particularly intriguing to enable substantial emission rate enhancement via the Purcell effect\cite{Happ2002,Riedrich-Moller2014}, which is primarily determined by the Q-factor to mode volume ratio. Therefore, PC cavities are used within diverse applications, such as enhancement of the internal quantum efficiency of single photon sources (SPS)\cite{Chang2006}, realization of compact high-resolution on-chip spectrometers\cite{Liapis2016}, photonic crystal lasers\cite{Loncar2002,Wu2004} and the generation of polaritons for explorations in cavity quantum electrodynamics\cite{Ohta2011}.

The main challenge of PC cavity fabrication lies in the periodic features, such as the holes, which need to scale with the target wavelength, yielding small lattice constants and consequently small hole radii for the UV-to-green spectral range. So far, numerous PC cavities have been developed covering the whole spectral range from VIS into UV using Si\cite{Akahane2003} and III-V semiconductors\cite{Chalcraft2007,Shih2007,Trivino2015,Neel2011,Arita2007,Choi2005,Meier2006}. For II-VI semiconductors, PCs have only been realised in ZnO-based heterostructures\cite{Wu2004,Hoffmann2016,Hoffmann2018} operating in the near-UV spectral range. However, to our knowledge, PC cavities have not yet been implemented in the selenides (e.g. ZnSe, MgSe, CdSe) which are able to bridge the spectral gap between the near-UV and the green spectral range.

Especially, ZnSe- and (Zn,Mg)Se-based heterostructures provide a large direct band gap and high oscillator strength. The minimal lattice mismatch between ZnSe and GaAs allows for nearly defect-free growth of ZnSe on standard GaAs substrates. More complex structures, implementing ZnSe QWs between ternary (Zn,Mg)Se barriers, enable even more intensive emission, smaller linewidths and the opportunity of spectral tuning by QW engineering\cite{Pawlis2011, Kutovyi2022}. Moreover, substitution of the ZnSe QW with one containing (Zn,Cd)Se also allows for accessing the full visible spectral range from blue to yellow\cite{Ruth2015, Finke2015}.

Besides these aspects, ZnSe and related materials have demonstrated enormous potential for modern quantum optical devices, such as indistinguishable single photon sources (SPS)\cite{Sanaka2009,Karasahin2022,Pettit2022,Sanaka2012} and optically controlled spin qubits\cite{Sleiter2013,Kim2014,Heisterkamp2015a}. Kutovyi et al.\cite{Kutovyi2022} demonstrated highly efficient nanopillar-based SPSs using the donor-bound exciton emission from Cl-donors in ZnSe QWs, where the external quantum efficiency was strongly enhanced by tailored solid immersion nanolenses on top of the nanopillars. Very recently, Jiang et al.\cite{Jiang2023} demonstrated Purcell enhancement of the emission from a ZnSe based SPS by fabrication of bullseye cavities. Further improvement of the external and internal quantum efficiency of such SPSs is feasible via both, Purcell enhancement and far-field tailoring when integrating the SPSs into PC cavities. Moreover, this approach will open up new perspectives for studies of exciton-photon coupling in ZnSe and ZnCdSe, which, until now, has been primarily investigated in one-dimensional (1D) polariton systems in ZnSe\cite{Pawlis2002,Curran2007,Sebald2012}.

In this work, we present the successful implementation of two-dimensional (2D) PC cavities in ZnSe QW heterostructures. Instead of adopting the common membrane design, the realization of which leads to strain relaxation and mechanical deformation in our material system, we chose a design with an underlying supporting layer, as similarly presented in  literature\cite{Phillips1999,Chow2000,Tokushima2000,Shih2007}. The pseudomorphically strained ZnSe/(Zn,Mg)Se heterostructures are grown on top of an AlAs buffer layer, which is post-growth wet oxidized into amorphous oxide. This approach offers not only mechanical stability and strain retainment, but also moderately high index guiding for optical confinement in the vertical direction.
The fabricated PC cavities were then subject to structural and elemental analysis to investigate the impact of the fabrication process. Optical characterization of the PC cavities using cross-polarized resonance scattering (RS) spectroscopy provides experimental evidence of the cavity modes. In order to refine the dielectric dispersion and to demonstrate the possibility of individual tunability of the cavity mode energy, we studied a large number of photonic crystals with varying dimensions and confirmed the experimental results by optical simulations using finite domain time difference (FDTD) method.
\section{Results and discussion}

\begin{figure}[!ht]
\makebox[\linewidth][c]{
    \centering
    \includegraphics{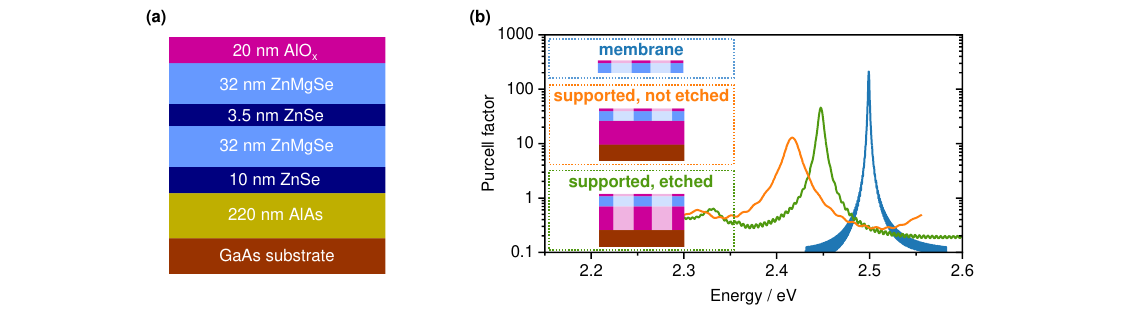}}
    \caption{(a) Scheme of the heterostructure layer stack from which the PC cavities are fabricated. (b) Comparing simulated Purcell factors for three different vertical designs (as sketched in the inset) using unmodified L3 PC cavity (a PC cavity design with three holes missing in the center).}
    \label{Fig1}
\end{figure}

We will first focus our discussions on a typical heterostructure, which is composed of a ZnSe QW enclosed in ZnMgSe barriers grown on top of an AlAs buffer layer. The overall layer stack is schematically shown in Fig.\ref{Fig1}a. First, using molecular beam epitaxy (MBE), a roughly \qty{220}{\nano\meter} thick AlAs buffer was grown on the GaAs-(001) substrate. After III-V growth, the sample was in-situ (maintaining continuous UHV conditions) transferred to the II-VI MBE chamber, in which the ZnSe/ZnMgSe heterostructure is grown. It consists of a \qty{10}{\nano\meter} thick ZnSe buffer and \qty{32}{\nano\meter} thick $\mathrm{Zn_{0.91}Mg_{0.09}Se}$ barriers enclosing a \qty{3.5}{\nano\meter} ZnSe quantum well (QW). Finally, the sample is further transferred into an atomic layer deposition (ALD) tool and capped with \qty{20}{\nano\meter} AlO\textsubscript{x}.
Note, that due to the larger lattice constants of ZnSe and ZnMgSe compared to GaAs, the overall II-VI heterostructures and also the AlAs buffer layer grow pseudomorphically on the GaAs substrate. The overall lattice relaxation of the quantum well is about \qty{2}{\percent} as confirmed by a XRD reciprocal space map shown in the supporting information (SI), section S1. This low value of strain relaxation indicates a negligible amount of misfit defects in the whole heterostructure.

In addition to the heterostructure, the design of the PC cavity also plays a crucial role in determining important parameters such as the Q-factor and mechanical stability. Therefore, we used the commercial FDTD solver \textsf{Lumerical} to perform FDTD simulations for three different vertical PC designs shown in the inset of Fig.\ref{Fig1}b, namely free-standing membrane structure, supported structure with intact underlying AlO\textsubscript{x} buffer and supported structure with the underlying AlO\textsubscript{x} buffer etched through.

The simulations were carried out by assuming a design of an unmodified L3 cavity (PC cavity with three holes missing in the center) with a lattice constant of \qty{175}{\nano\meter} and a radii ratio (i.e. radius/lattice constant) of 0.26. For the purpose of comparison, we restrict ourselves to the selection of a single specific mode (Mode-1, see Fig.\ref{Fig4}a for details). In the simulation model, we assumed for the overall II-VI semiconductor heterostructure the same refractive index dispersion, e.g. as it would be composed of a \qty{72}{\nano\meter} thick ZnMgSe layer. This gives an adequate fit of the overall heterostructure model with ZnSe buffer and ZnMgSe barriers obtained from ellipsometry measurements. Since there are no experimentally systematic data available for the dispersion relation of arbitrary ZnMgSe compounds, we determined the refractive index $n(E)$ using the equations for a single-effective-oscillator (SEO) model from reference\cite{Liu2004}:
\begin{equation}
\begin{aligned}
 n\left(E\right) = \sqrt{1+\frac{F_0{E_0}^2}{{E_0}^2-E^2}+\frac{0.005{E_g}^2({E_g}^2-{E}^2)}{({E_g}^2-{E}^2)^2+0.0004E^2}}\quad\mathrm{(energy\;unit:eV)}\\
 F_0 = -0.281x^2-0.259x+4.899; \;
 E_0 = -0.385x^2+1.524x+5.207
  \label{eqn1}
\end{aligned}
\end{equation}
with $F_0$ and $E_0$ representing the effective oscillator strength and band gap of the SEO model, respectively. We only modified the third term in the square root into a Lorentz oscillator to avoid divergence near the ZnMgSe band gap energy $E_g$. At low temperatures, we assumed $E_g = 2.82 +1.28x$ (\unit{\electronvolt}) as reported in Ref.\cite{Puls1998}, where $x$ represents the Mg concentration. The imaginary part of the dispersion relation is automatically generated in the FDTD solver via Kramers-Kronig relation and is negligible except for the photon energies close to the ZnMgSe band gap energy. More details can be found in section S2 of the supporting information.

For the dispersion relation of the wet-oxidized former AlAs-buffer layer (and also for the AlO\textsubscript{x} capping layer), we used values from Tauc-Lorentz model, as this model provides the most accurate fit for the heterostructures according to ellipsometry measurements. The obtained refractive index (see SI section S2) is close to that reported by Ref.\cite{Hirai2012,Bek1999} for wet-oxidized former AlAs into amorphous oxide and considering a negligible absorption coefficient in this material. At a wavelength of \qty{440}{\nano\meter}, the corresponding refractive indices of $\mathrm{Zn_{0.91}Mg_{0.09}Se}$, oxide buffer and GaAs\cite{Palik1997} are \num{2.78}, \num{1.66} and \complexnum{5.00 + 1.07i} respectively, forming index guiding in the vertical direction in case of the supported design with intact oxide buffer below, though weaker than that of the standard membrane design.

\begin{table}[H]
  \caption{Simulation results for the three different vertical designs}
  \label{tbl}
  \begin{tabular}{ccccc}
    \hline
    Design  & $E_{mode}$ / \unit{\electronvolt} & $Q$ & $V_{mode}$ / \unit{\cubic\nano\meter} & $F_P$\textsuperscript{\emph{*}}  \\
    \hline
    membrane  & \num{2.499} & \num{1800} & \num{4.0e6} & \num{210}\\
    supported, not etched & \num{2.417} & \num{140} & \num{6.2e6} & \num{13}\\
    supported, etched  & \num{2.447} & \num{410} & \num{4.7e6} & \num{46}\\
    \hline
  \end{tabular}
  \\\textsuperscript{\emph{*}} Purcell factor is calculated with the formula $F_p = \frac{3}{4\pi^2}\frac{Q}{V_{mode}}{\left(\frac{\lambda}{n}\right)}^3 $, which depicts the Purcell enhancement averaged for the effective cavity volume, independent on where our dipole source is located. Here in the formula, $\lambda$ is the cavity mode wavelength and  n is assumed to be the refractive index of ZnMgSe at $\lambda$.
\end{table}

The simulation results are shown in Fig.\ref{Fig1}b as well as in Table \ref{tbl}. The membrane structure provides the strongest index guiding of all three designs, which is verified by the largest Q-factor of \num{1800}. However, the fabrication of such membrane design from our layer stack is experimentally challenging and hardly feasible. Under-etching the material below the ZnSe/ZnMgSe heterostructure induces relaxation of the residual compressive strain in the overall layer stack, which leads to the formation of extended defects and cracks in the PC structure. On the other hand, the supported design with full oxide buffer provides good mechanical stability, but suffers from poor index guiding, causing radiation losses into the substrate. As a result, a Q-factor of only \num{140} is achieved in this case. Aiming for a compromise between mechanical stability and vertical optical confinement, we opt for a design, in which the holes of the photonic crystals are also etched through the underlying oxide buffer. The etching reduces the effective refractive index and therefore enhances the optical confinement in the II-VI layers, which results in an increased Q-factor of \num{410}. For this specific vertical design, further optimization in the horizontal L3 cavity design by modification of the three inner hole pair positions, reveals an additional roughly two-fold increase of the Q-factor. The effect of such enhancement is significantly smaller than that obtained in membrane PC cavities, as in this specific vertical design, the limiting factor of optical loss is mainly vertical leakage instead of end-hole scattering. Therefore in the current investigations, we confine our focus to the unmodified L3 cavity as our preferred PC cavity design.

Having established the PC cavity designs, we now turn our attention to the fabrication of the PC cavities. The device fabrication process employs a top-down technique, shortly summarized in the following (see SI section S3 for details). First, \qty{40}{\nano\meter} Cr is deposited on the pristine layer stack as a hard mask. Secondly, the PC cavity design is defined using e-beam lithography. The pattern is then transferred into the Cr layer using reactive ion etching (RIE) with a gas combination of Cl\textsubscript{2}/O\textsubscript{2}. Next, several successive etching steps are performed, starting with wet-chemical etching of the AlO\textsubscript{x} capping layer using developer \textsf{AZMIF326}. Then RIE dry etching with H\textsubscript{2}/Ar and Cl\textsubscript{2}/Ar/CH\textsubscript{4} chemistry is applied for etching through the ZnSe/ZnMgSe heterostructure and the AlAs layer, respectively. Subsequently, the sample undergoes wet oxidation in a reaction chamber at \qty{350}{\celsius} using water vapor carried by nitrogen gas flow. After wet oxidation, the \qty{220}{\nano\meter} thick AlAs buffer is transformed into roughly \qty{190}{\nano\meter} thick AlO\textsubscript{x} (shrinkage ratio of \qty{86}{\percent}, similar to reported values\cite{MacDougal1994,Takamori1996}), where the shrinkage is validated by means of ellipsometry measurements and scanning electron microscopy (SEM). Finally, the Cr hard mask is removed by subsequent utilization of O\textsubscript{2} and Cl\textsubscript{2}/O\textsubscript{2} plasma.

\begin{figure}[!ht]
\makebox[\linewidth][c]{
    \centering
    \includegraphics{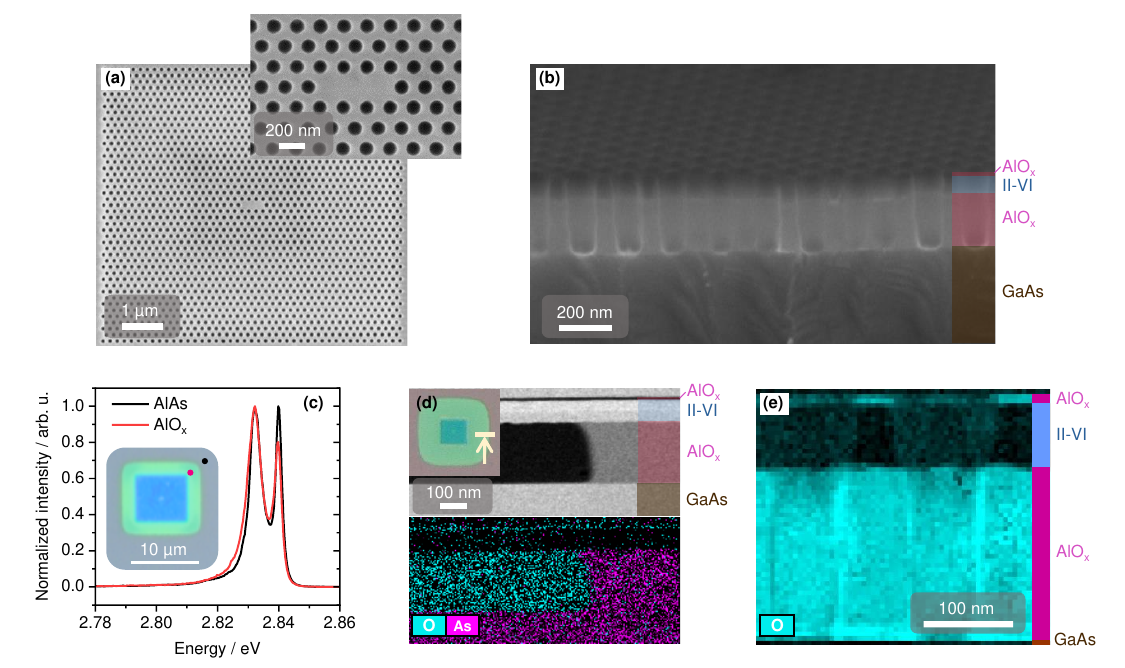}
    }
    \caption{SEM and TEM characterization of a typical fabricated sample. (a) Top-view SEM micrograph of a PC cavity after processing. The zoomed-in inset shows the cavity region. (b) Cross-sectional SEM micrograph of a PC cavity of the same design revealing the clear interface boundaries between the stack materials.  (c) PL spectra of the pristine region of the sample (black spot in the inset) with AlAs below, and of the wet-oxidized region of the sample (red spot) with AlO\textsubscript{x} layer below the II-VI material, respectively. The inset shows the optical micrograph around one typical PC cavity. (d) Cross-sectional HAADF micrograph at the wet oxidation front and the corresponding STEM-EDS elemental mapping of arsenic (purple) and oxygen (cyan). One typical oxidation front is marked in the optical micrograph around one PC cavity in the inset. (e) STEM-EELS elemental mapping revealing the presence of oxygen at the cross-section of a PC structure. }
    \label{Fig2}
\end{figure}

Following the fabrication process, Fig.\ref{Fig2}a presents a top-view SEM image of a typical PC cavity. By fitting circle sizes and center positions of a zoom-in micrograph, we determine the average hole diameter $d$, average lattice constant $L$ and therefore the average radii ratio $R = d / 2L$ for each individual PC cavity, which is later optically characterized (see SI section S4 for details of the fitting process). For the PC cavity shown in Fig.\ref{Fig2}a, the corresponding parameters are \qty{102+-4}{\nano\meter}, \qty{177+-4}{\nano\meter} and \num{0.289+-0.013}. A cross-sectional SEM micrograph of the PC cavity, shown in Fig.\ref{Fig2}b, demonstrates nearly vertical sidewalls and a good anisotropy of the applied etching processes.

To investigate the influence of wet oxidation on the sample's strain condition, we measured photoluminescence (PL) spectra from the same sample at untreated and wet-oxidized regions, as indicated by the black and red spots in the inset of Fig.\ref{Fig2}c, respectively. Note, that the sizes of the spots correspond to the area selection determined by a \qty{100}{\micro\meter}-diameter pinhole. The measurement was carried out at \qty{10}{K} under above-band excitation with a \qty{377}{\nano\meter} continuous-wave laser and with a spectrometer equipped with a \qty{600}{lines\per\milli\meter} blazed grating. The PL spectra in Fig.\ref{Fig2}c reveal a two-peak structure in the excitonic transition region of the ZnSe QW, where the peaks at \qty{2.840}{\electronvolt} and \qty{2.832}{\electronvolt}, respectively, correspond to the emission from free excitons and trions in the QW. The trion peak has a comparable intensity as the free exciton peak, which is most likely caused by residual n-type background doping in our growth chamber. Remarkably, the PL peak positions shown in Fig.\ref{Fig2}c remain at the same spectral energy for both, untreated and wet-oxidized regions, which indicates complete preservation of the compressive strain after the wet oxidation process.

Furthermore, for a better understanding of the influence of the fabrication process on the semiconductor properties, we conducted scanning transmission electron microscopy (STEM) analysis of a typical PC cavity after fabrication. For analysis preparation, two lamellas across a PC structure and across a wet oxidation front in the sample are lifted out using focused ion beam (FIB) milling. Fig.\ref{Fig2}d shows the cross-sectional high-angle annular dark-field (HAADF) micrograph  of the wet oxidation front and the corresponding elemental maps of arsenic and oxygen obtained by energy dispersive spectroscopy (EDS). From the change in contrast across the middle layer shown in the HAADF micrograph, the material changes across the oxidation front. Since arsenic is not found in the EDS map on the oxidized side of the oxidation front, the transformation of AlAs into amorphous oxide (most probably aluminium hydroxide\cite{Li2007,Hirai2012}) is confirmed. For convenience, we refer to this oxide layer as AlO\textsubscript{x} in the following discussion. Fig.\ref{Fig2}e shows the elemental mapping of oxygen by electron energy loss spectroscopy (EELS) at the cross-section of the PC structure. The sidewalls of the etched holes in the ZnSe/ZnMgSe layers show the presence of oxygen, indicating a certain degree of ZnMgSe oxidation. This is a side-effect that results from the wet oxidation and the Cr-removal steps with O\textsubscript{2} and Cl\textsubscript{2}/O\textsubscript{2} plasma. This oxidation impacts the refractive index of the material and, therefore, alters the cavity mode energy and Q-factor of the fabricated PC cavities. As discussed in the following analysis, this effect is taken into account when simulating the PC cavities.

\begin{figure}[!ht]
\makebox[\linewidth][c]{
    \centering
    \includegraphics{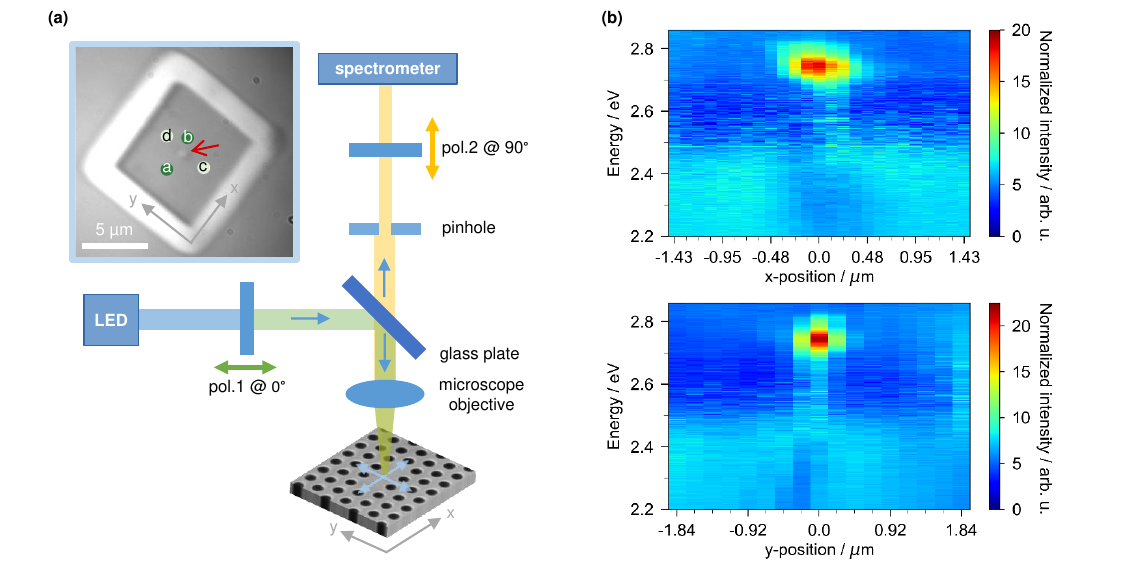}
    }
    \caption{(a) Schematic drawing of the cross-polarized RS spectroscopy setup. The inset shows the spatial-dependent reflection map of a PC cavity subject to x- and y- line scans, with the starting/ending scanning positions marked as "a"/"b"  and "c"/"d" inside the circles (with the diameter of the circles corresponding to the spatial resolution obtained with our pinhole). The cavity region is marked with a red arrow. (b) Normalized cross-polarized RS spectra of x- and y- line scans.}
    \label{Fig3}
\end{figure}

Following our PC cavity design and sample preparation, the subsequent optical characterization of the PC cavities is conducted using a cross-polarized resonance scattering (RS) spectroscopy setup\cite{McCutcheon2005,Liapis2016}, as schematically illustrated in Fig.\ref{Fig3}a. In this setup, excitation is performed using a white light LED and the first polarizer sets the polarization at a \ang{45} angle tilt to the cavity direction (long axis of L3 defect) and in the sample plane. On one hand, this ensures excitation and collection of the cavity modes, which are polarized either parallel or perpendicular to the cavity axis (as discussed below in Fig.\ref{Fig4}a). On the other hand, a second polarizer, rotated by \ang{90} relative to the first one, ensures that the directly reflected light from the sample surface is filtered out. 
In addition, we use a microscope objective with a large numerical aperture of \num{0.9} to attain a large collection angle, which circumvents the complexities associated with the far-field projection of various cavity modes. A pinhole with a diameter of \qty{100}{\micro\meter} is implemented as a spatial filter for the region of interest and a blazed grating with  \qty{150}{lines\per\milli\meter} is chosen in the spectrometer to resolve a large spectral range covering the cavity modes of various PC cavity designs. The experiments are carried out at a temperature of \qty{10}{\kelvin}.

In order to verify the presence of cavity modes, we first performed line scans in x- and y-direction across a PC cavity (with dimensions L=\qty{186+-4}{\nano\meter}, R=\num{0.308+-0.016}). A reflection map of the investigated PC cavity is shown as inset in the left upper corner of Fig.\ref{Fig3}a, where the starting/ending positions of the scans are labeled as "a"/"b" and "c"/"d" inside the circles and the circle size represents the spatial resolution obtained with our pinhole. 
The results of the line scans in both directions are shown in Fig.\ref{Fig3}b, with the obtained RS spectra normalized to one reference spectrum taken far away from the cavity region. In the center of the cavity region, a resonant signal is visible, which we attribute to the presence of specific cavity modes at an energy of \qty{2.727}{\electronvolt}.

\begin{figure}[!ht]
\makebox[\linewidth][c]{
    \centering
    \includegraphics{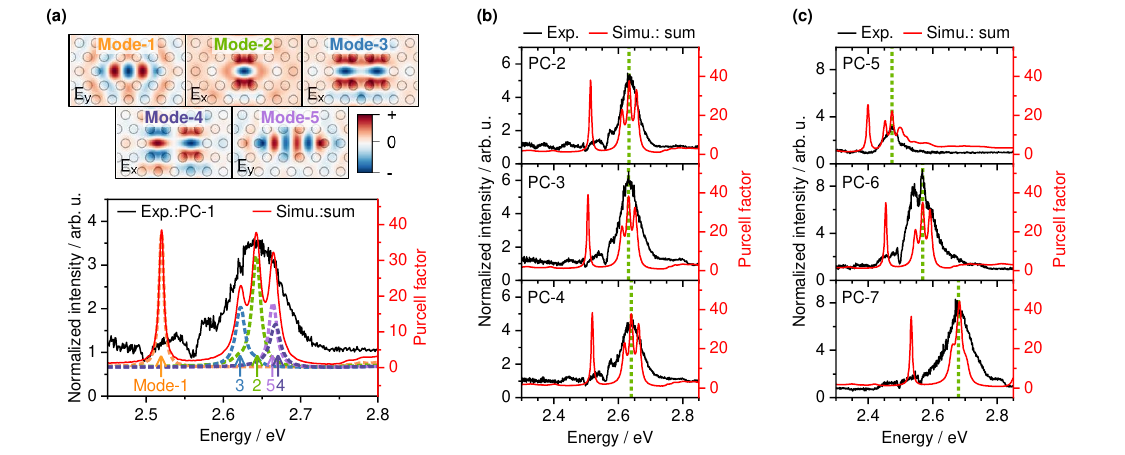}
    }
    \caption{Normalized cross-polarized RS spectra of several measured PC cavities and the corresponding simulated sum spectra of Purcell factor. For the structure PC-1, the mode profile of the dominant electric field and the individual contribution of Purcell factor from each present mode are shown.  The structures are: (a) PC-1 with L=\qty{181+-4}{\nano\meter} and R=\num{0.274+-0.013} (b) Three PC cavities with similar parameters, PC-2/ PC-3/ PC-4: L=\qty{181+-3}{\nano\meter}/ \qty{181+-4}{\nano\meter}/ \qty{181+-4}{\nano\meter}, R=\num{0.269+-0.014}/ \num{0.276+-0.017}/ \num{0.272+-0.017}. (c) Three PC cavities with similar lattice constants but different radii ratios, PC-5/ PC-6/ PC-7: L=\qty{186+-5}{\nano\meter}/ \qty{186+-3}{\nano\meter}/ \qty{185+-3}{\nano\meter}, R=\num{0.223+-0.020}/ \num{0.264+-0.014}/ \num{0.303+-0.011}. The mode energy of the dominant Mode-2 is marked with the green dashed lines in (b) and (c).}
    \label{Fig4}
\end{figure}

For a specific PC cavity, detailed mode analysis from simulation was carried out. Fig.\ref{Fig4}a depicts the background-normalized RS spectrum of a single PC cavity (L=\qty{181+-4}{\nano\meter} and R=\num{0.274+-0.013}) together with the simulated mode spectra and Purcell factors of different modes (colored dashed spectra), as well as the overall response given by the sum of all modes (red spectrum). Altogether five modes are revealed for this particular PC cavity in the investigated spectral region. The electric field distribution of each mode is also shown in Fig.\ref{Fig4}a, where $E_y$ is the dominant electric field for Mode-1 and Mode-5 and $E_x$ that for the other modes, respectively. The electric field of Mode-1 is strongly concentrated in the cavity region, unlike that of the other modes, which is more delocalized and penetrates into the neighboring air holes. This observation accounts for the lowest cavity mode energy of Mode-1 in the simulation. Further theoretical investigation of the far-field projection reveals that each mode is polarized with a polarization angle aligned with the dominant electric field direction, which renders the measurement with the above-described cross-polarized RS spectroscopy setup feasible. For the simulation, the ZnMgSe dispersion relation was scaled by a factor of \qty{93+-1}{\percent} (see SI section S2 for details), so that the simulated peak energies (colored spectra in Fig.\ref{Fig4}a) match the experimental one (black spectrum). This modification is justified under consideration of the process-related oxidation of ZnMgSe into ZnMgSeO as previously discussed (following the STEM-EELS results in Fig.\ref{Fig2}e), which would reduce the overall effective refractive index in the structure. Additionally in our observation, the broad RS signal region around \qty{2.65}{\electronvolt} can be attributed to the superposition of a cluster of cavity modes Mode-2-5, as underlined by the simulations. The possible absence of Mode-1 in the experimental spectrum might be related to an ineffective coupling between the electronic states of the QW exciton and the y-polarized photonic modes. However, further investigation of this specific behaviour of Mode-1 is beyond the scope of this paper.

To address the influence of fabrication tolerances and microscopic PC designs on the experimentally observed resonant signals, normalized experimental spectra of six PC cavities are shown in Fig.\ref{Fig4}b and Fig.\ref{Fig4}c, alongside their corresponding simulated cavity response and Purcell factors. The best agreement between simulation and experimental results was again found by using the previously determined refractive index scaling factor of \qty{93+-1}{\percent} of ZnMgSeO. For all six investigated cavities, Mode-2 (its energy marked by a green dashed line) is the dominant mode in the cluster.
Taking a closer look, Fig.\ref{Fig4}b shows the RS spectra of three PC cavities (PC-2, PC-3 and PC-4) of the same design but with slight parameter variations, which stem from fabrication tolerances (lattice constant L=\qty{181+-3}{\nano\meter}, \qty{181+-4}{\nano\meter}, \qty{181+-4}{\nano\meter} and radii ratio R=\num{0.269+-0.014}, \num{0.276+-0.017}, \num{0.272+-0.017} as determined from the respective SEM micrographs, where the obtained lattice constants closely match the intended design of \qty{180}{\nano\meter}). The experimental cavity mode energies are extracted from the peak positions and their values vary only slightly (\qtylist{2.629;2.632;2.646}{\electronvolt}), which confirms substantial robustness of the cavity mode energy against fabrication tolerances. On the other hand, Fig.\ref{Fig4}b shows spectra of three PC cavities (PC-5, PC-6 and PC-7) selected from different designs with nearly equal lattice constants close to the designed value of \qty{185}{\nano\meter}, but having distinct radii ratios (L=\qty{186+-5}{\nano\meter}, \qty{186+-3}{\nano\meter}, \qty{185+-3}{\nano\meter}, R=\num{0.223+-0.020}, \num{0.264+-0.014}, \num{0.303+-0.011}). The experimentally observed cavity mode energy significantly blue-shifts with increasing hole radius (\qtylist{2.474;2.568;2.670}{\electronvolt}), as expected from the corresponding simulations. This result demonstrates the possibility of tuning the cavity mode energy over a substantially large spectral range, which in the future can be applied to bring the cavity mode close to resonance with the QW emission for Purcell enhancement.

\begin{figure}[!ht]
\makebox[\linewidth][c]{
    \centering
    \includegraphics{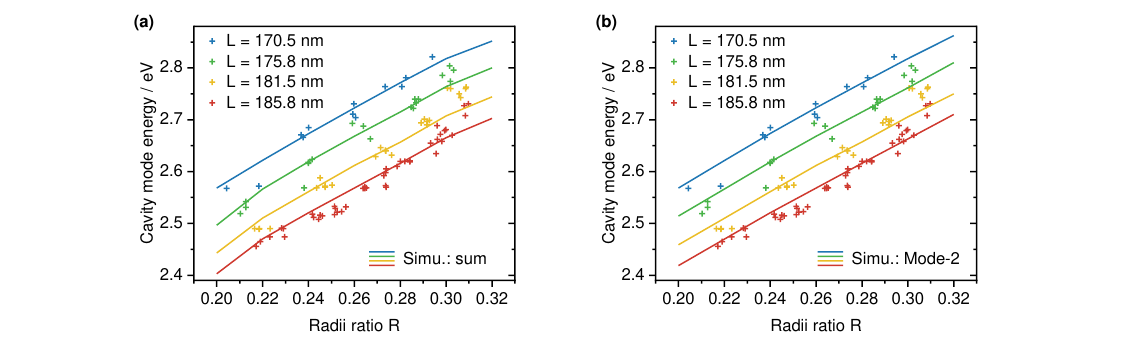}}
    \caption{Cavity mode energies extracted from experimental spectra of various PC cavities. (a) Cluster peak energies obtained by summing up the mode profiles of Mode-2-5 versus radii ratio. (b) Peak energy of the dominant Mode-2 versus radii ratio. Dots correspond to the experimentally measured energies and lines are obtained from the theoretical simulations using the corresponding lattice constants labeled in the legend of the diagrams. }
    \label{Fig5}
\end{figure}

In order to gain a more comprehensive insight into the PC cavities and especially to quantify the dispersion relation in our heterostructures including the abovementioned formation of ZnMgSeO, we extensively investigated nearly 100 individual PC cavities with varying lattice constants and radii ratios and extracted the experimental cavity mode energies from the measured RS spectra. In Fig.\ref{Fig5}a and b, the experimentally determined mode energies are plotted versus the radii ratios (dots). For each measured data point, the real radii ratio was individually determined by extraction of the average diameter of the holes and their mean distance from SEM micrographs of the investigated PC cavities. All obtained datasets were assigned to four main groups of structures with different designed lattice constants. Cavity mode simulations were performed using the average value of the determined lattice constants for each group (\qty{170.5+-0.4}{\nano\meter} (blue), \qty{175.8+-0.3}{\nano\meter} (green), \qty{181.5+-0.4}{\nano\meter} (yellow), \qty{185.8+-0.5}{\nano\meter} (red)) and with radii ratios ranging from \numrange{0.20}{0.32}.
Experimental data points and simulation results are compared in Fig.\ref{Fig5}a with respect to the cluster peak energy as obtained by summing up the mode profiles of Mode-2-5 and in Fig.\ref{Fig5}b, when only the spectral energy of the dominant Mode-2 is considered, respectively. In both cases, the experimental data are in good agreement with the simulated dependence of the radii ratio. Only a small deviation can be seen in the case of radii ratios larger than 0.30, which could be due to slight underestimation of the degree of ZnMgSe oxidation for large radii ratios in the model. In this case, the smaller volume of ZnMgSe material between the air holes may ease the oxidation process, requiring an additional correction of the scaling factor for the ZnMgSe dispersion relation.

The observation of cavity modes, which spans a broad range of cavity mode energy from \qty{2.456}{\electronvolt} to \qty{2.821}{\electronvolt}, proves the successful implementation of 2D PC cavities in ZnSe QW heterostructures. At the same time, the excellent agreement between our simulation and experimental results demonstrates the pertinence of our simulation model using our quantified dispersion relation of ZnMgSeO.
\section{Conclusions}
In this work, we demonstrate the successful implementation of 2D photonic crystal cavities within ZnSe QW heterostructures as tailored for the blue/green spectral range. To the best of our knowledge, no prior investigation of 2D PC cavities has been reported in this specific material system. In our innovative approach, the QW heterostructure is grown on top of an AlAs buffer layer, which then is post-growth wet-oxidized into AlO\textsubscript{x}. This technique is a mandatory requirement to provide excellent mechanical stability as well as sufficient optical confinement in the vertical direction via index guiding.

Detailed analysis using scanning electron microscopy revealed excellent structural properties of the fabricated PC cavities, with steep side walls of the etched holes signifying remarkable etching anisotropy. This is particularly noteworthy considering the small characteristic length scale, represented by etched holes with diameters in the range of \qtyrange{70}{115}{\nano\meter}. The applied elemental mapping techniques, namely energy dispersive spectroscopy and electron energy loss spectroscopy, not only confirmed the complete wet-oxidation of AlAs into AlO\textsubscript{x}, but also revealed partial oxidation of the II-VI heterostructure. In particular, oxidation of the ZnMgSe barriers leads to the formation of ZnMgSeO and alters the corresponding dispersion relation.
The optical characterization of fabricated PC cavities using cross-polarized resonance scattering spectroscopy verified the existence of cavity modes in the central PC cavity region, of which the energy can be tuned over a broad spectral range of \qtyrange{2.456}{2.821}{\electronvolt}, depending on the microscopic PC design. Moreover, we quantified the effective dispersion relation of the partially oxidized QW heterostructure and determined a scaling factor of \qty{93+-1}{\percent} by fitting the measured cavity mode energies with simulation results performed with a FDTD solver.

The obtained results represent a solid experimental and theoretical basis for further investigations of PC cavities in the II-VI semiconductor material system. Those may cover for example interaction between cavity and QW emission, the enhancement of Purcell factors or tailoring of the angular far-field projection of the cavity modes.
\begin{acknowledgement}
The authors gratefully acknowledge the technical support by the staff of the Helmholtz Nano Facility (HNF) of Forschungszentrum Jülich in the fabrication process as well as the support from our colleagues.
This work is supported by Deutsche Forschungsgemeinschaft (DFG, German Research Foundation) under Germany’s Excellence Strategy - Cluster of Excellence: Matter and Light for Quantum Computing (ML4Q) (EXC 2004 1-390534769) X.C. and J.L. gratefully acknowledge support from the Air Force Office of Scientific Research (FA9550-20-0066).
\end{acknowledgement}
\newpage
\newpage
\bibliography{6-references}
%\end{multicols}

\end{document}